\iftrue
\documentclass[aps,prl,twocolumn,
			   groupedaddress,superscriptaddress,
			   amsfonts,amssymb,amsmath,
			   citeautoscript,
			   a4paper]{revtex4-1}
\else
\documentclass[aps,pra,preprint,
			   groupedaddress,superscriptaddress,
			   amsfonts,amssymb,amsmath,
			   citeautoscript,
			   a4paper]{revtex4-1}
\fi

\usepackage[pdftex]{hyperref}
\hypersetup{colorlinks,
			linkcolor={blue!75!black!80!yellow},
			citecolor={blue!75!black!80!yellow},
			urlcolor={blue!75!black!80!yellow},
			pdfstartview=FitH}
\usepackage{graphicx}
\usepackage{mathrsfs}
\usepackage{xspace}
\usepackage{braket}
\usepackage{xr}
\usepackage{gensymb} % to use \degree command
\usepackage{xcite}
\usepackage{xcolor,soul}
\usepackage{stmaryrd} % provides \llbracket and \rrbracket for difference-operations
\usepackage[UKenglish]{babel}
\usepackage{placeins} % provides \FloatBarrier
\usepackage{bbm}
\usepackage{physics}

\usepackage{siunitx}
\DeclareSIUnit[number-unit-product=]\percent{\char`\%} % remove space before percentage "units"

\usepackage{txfonts}  %Times Roman fonts
\usepackage{txfontsb} %Addition for txfonts, including old style numerals and greek

\makeatletter
\newcommand*{\addFileDependency}[1]{% argument=file name and extension
  \typeout{(#1)}% latexmk will find this if $recorder=0 (however, in that case, it will ignore #1 if it is a .aux or .pdf file etc and it exists! if it doesn't exist, it will appear in the list of dependents regardless)
  \@addtofilelist{#1}% if you want it to appear in \listfiles, not really necessary and latexmk doesn't use this
  \IfFileExists{#1}{}{\typeout{No file #1.}}% latexmk will find this message if #1 doesn't exist (yet)
}
\makeatother

\newcommand*{\myexternaldocument}[1]{%
    \externaldocument{#1}%
    \addFileDependency{#1.tex}%
    \addFileDependency{#1.aux}%
}

\makeatletter
\renewcommand\@make@capt@title[2]{%
	\@ifx@empty\float@link{\@firstofone}{\expandafter\href\expandafter{\float@link}}%
	\sffamily{\textbf{#1}}\@caption@fignum@sep#2
}%

\makeatother

\newcommand{\iu}{\mathrm{i}}

\newcommand{\appropto}{\mathrel{\vcenter{
			\offinterlineskip\halign{\hfil$##$\cr
				\propto\cr\noalign{\kern2pt}\sim\cr\noalign{\kern-2pt}}}}}

\newcommand{\gqu}{g_\mathrm{Q}}
\newcommand{\gqupoint}{\abs{\gqu}^2_\mathrm{3D}}
\newcommand{\gqusheet}{\abs{\gqu}^2_\mathrm{2D}}
\newcommand{\Tmatrix}{\mathbbm{T}}

\newcommand{\Ti}{\mathbbm{T}_{\mathit{i}}}

\newcommand{\chargedensity}{\mathit{q}}

\newcommand{\Einc}{\mathbf{E}_{\rm{inc}}}

\newcommand{\einc}{\mathbf{e}_{\rm{inc}}}
\newcommand{\polfield}{\mathbf{P}}
\newcommand{\Powerext}{\textit{P}_{\rm{ext}}}

\newcommand{\kvmax}{\textit{k}_{v,0}}
\newcommand{\kmax}{\textit{k}_{0}}
\newcommand{\omegam}{\omega_0}
\newcommand{\omegap}{\omega_{\rm{p}}}

\newcommand{\kapparhomax}{\kappa_{\rho,0}}
\newcommand{\BesselKzero}{\textit{K}_0}
\newcommand{\BesselKfirst}{\textit{K}_1}

\newcommand{\rv}{\mathbf{r}}
\newcommand{\smatrix}{\hat{\mathit{S}}}

\newcommand{\Bi}{\mathbbm{E}_\textit{i}}

\newcommand{\econstant}{\tau}

\usepackage{textcomp} % for \textrightarrow
\usepackage{xifthen}
\usepackage{etoolbox}
\newboolean{togglecomments}
\newboolean{togglechanges}

\setboolean{togglecomments}{true}
\setboolean{togglechanges}{false}

\newcommand{\comment}[2]{%
    \ifbool{togglecomments}%
    {\textcolor{blue!70!black}{\small\textsf{%
    \textsuperscript{\textsc{\textsf{\MakeLowercase{#1}}}}%
    [#2]}}} % if true, show comments
    {}}     % if false, do nothing
\newcommand{\swap}[2]{\ifbool{togglechanges}
    {#2}  % updates-only version
    {\textcolor{red!70!black}{[#1]}\textrightarrow{}\textcolor{green!50!black}{[#2]}}}
\newcommand{\remove}[1]{\ifbool{togglechanges}
    {}    % updates-only version
    {\textcolor{red!70!black}{#1}}}
\newcommand{\inset}[1]{\ifbool{togglechanges}
    {#1}  % updates-only version
    {\textcolor{green!50!black}{#1}}}

\newcommand{\optional}[1]{\ifbool{togglechanges}
    {}    % (minimal) updates-only version
    {\textcolor{yellow!50!orange!80!gray}{#1}}}

\newcommand{\citeremind}[1]{%
    [\textcolor{blue!75!black!80!yellow}{
        $\blacksquare$%
	    \ifthenelse{\isempty{#1}}
	        {}
	        {\textsuperscript{\tiny\textsf{#1}}}%
	}]\xspace}

\newcommand{\eg}{e.g.\@\xspace}

\newcommand{\hkuaffil}{\footnotesize Department of Physics and HK Institute of Quantum Science and Technology,
The University of Hong Kong, Pokfulam, Hong Kong, China}
\newcommand{\Yaleaffil}{\footnotesize Department of Applied Physics \& Energy Sciences Institute, Yale University, New Haven, CT 06520, USA}
\newcommand{\zjuaffil}{\footnotesize
Interdisciplinary Center for Quantum Information, State Key Laboratory of Modern Optical Instrumentation, College of Information Science and
Electronic Engineering, Zhejiang University, Hangzhou 310027, China}
\newcommand{\Technionaffil}{\footnotesize Solid State Institute and Faculty of Electrical and Computer Engineering, Technion – Israel Institute of Technology, 32000 Haifa, Israel}

\myexternaldocument{SM}

\begin{document}

%-----TITLE-----
\title{
Maximal quantum interaction between free electrons and photons
}

%-----AUTHORS AND AFFILIATIONS-----
\author{Zetao~Xie}
\thanks{Z.~X. and Z.~C. contributed equally to this work.}
\affiliation{\hkuaffil}
\author{Zeling~Chen}
\thanks{Z.~X. and Z.~C. contributed equally to this work.}
\affiliation{\hkuaffil}
\author{Hao~Li}
\affiliation{\Yaleaffil}
\author{Qinghui~Yan}
\affiliation{\zjuaffil}\affiliation{\Technionaffil}
\author{Hongsheng~Chen}
\affiliation{\zjuaffil}
\author{Xiao~Lin}
\affiliation{\zjuaffil}
\author{Ido~Kaminer}
\affiliation{\Technionaffil}
\author{Owen~D.~Miller}
\email{owen.miller@yale.edu}
\affiliation{\Yaleaffil}
\author{Yi~Yang}
\email{yiyg@hku.hk}
\affiliation{\hkuaffil}

\begin{abstract}

The emerging field of free-electron quantum optics enables electron-photon entanglement and holds the potential for generating nontrivial photon states for quantum information processing. 
Although recent experimental studies have entered the quantum regime, rapid theoretical developments predict that qualitatively unique phenomena only emerge beyond a certain interaction strength.
It is thus pertinent to identify the maximal electron-photon interaction strength and the materials, geometries, and particle energies that enable one to approach it.
We derive an upper limit to the quantum vacuum interaction strength between free electrons and single-mode photons, which illuminates the conditions for the strongest interaction.
Crucially, we obtain an explicit energy selection recipe for electrons and photons to achieve maximal interaction at arbitrary separations and identify two optimal regimes favoring either fast or slow electrons over those with intermediate velocities.
We validate the limit by analytical and numerical calculations on canonical geometries and provide near-optimal designs indicating the feasibility of strong quantum interactions.
Our findings offer fundamental intuition for maximizing the quantum interaction between free electrons and photons and provide practical design rules for future experiments on electron-photon and electron-mediated photon-photon entanglement. 
They should also enable the evaluation of key metrics for applications such as the maximum power of free-electron radiation sources and the maximum acceleration gradient of dielectric laser accelerators.

\end{abstract}

\maketitle

Free--electron--light interaction with structured optical environments leads to photon emission through diverse radiation mechanisms, including Cherenkov radiation, Smith-Purcell radiation, transition radiation, and incoherent cathodoluminescence~\cite{de2010optical,polman2019electron,garcia2021optical,talebi2018electron,rivera2020light,shiloh2022miniature,roques2023free}.
These interaction processes lie at the heart of modern electron microscopy~\cite{de2010optical,polman2019electron,garcia2021optical}, spectroscopy~\cite{varkentina2022cathodoluminescence}, radiation sources~\cite{massuda2018smith,roques2019towards,yang2023photonic,gong2023interfacial}, particle detection~\cite{lin2018controlling,lin2021brewster}, free--electron laser~\cite{pellegrini2016physics}, accelerators~\cite{sapra2020chip,shiloh2021electron,fishman2023imaging,chlouba2023coherent}, and biomedical imaging~\cite{roques2022framework,kurman2023purcell}.
Recent theoretical advances have shown that the quantum-photonic nature of light could become critical for describing the interaction~\cite{kfir2019entanglements,di2019probing}.
It was further predicted that the interaction could shape and entangle electron and photon wavefunctions~\cite{bendana2011single,kfir2019entanglements,di2020free,ben2021shaping,di2021modulation,kfir2021optical,karnieli2021coherence,konevcna2022entangling}, and induce entanglement among electrons~\cite{kfir2019entanglements} and photons~\cite{baranes2022free} themselves.
The interaction has also been proposed for free-electronic topological probes and quantum simulators~\cite{peng2019probing,pan2022synthetic,li2024topologically}.
Such a wide range of predicted quantum-optical interactions now form the basis of free-electron quantum optics.

With the rapid developments of photon-induced near-field electron microscopy (PINEM)~\cite{barwick2009photon,park2010photon,garcia2010multiphoton} on photonic platforms~\cite{kfir2020controlling,wang2020coherent,henke2021integrated,yang2024free}, the predicted quantum free-electron-light interaction has been realized in a few experiments~\cite{dahan2021imprinting,adiv2023observation,feist2022cavity}.
The crucial parameters for describing these interactions are the dimensionless spontaneous coupling strength $\gqu$ with vacuum and its stimulated counterpart $g$ with external optical fields ($\gqu$ and $g$ are sometimes respectively denoted as $\beta_0$ and $\beta$ in the literature, \eg Ref.~\cite{vanacore2018attosecond,polman2019electron,di2019probing}). 
$\gqu$ can be calculated via the spectral integration of spontaneous electron energy loss~\cite{kfir2019entanglements,di2019probing,kfir2021optical,di2021modulation,huang2023electron,baranes2023free},
and the two parameters relate to each other via $g = \sqrt{N}\gqu$ where $N$ is the number of photons injected into the optical mode from a pump laser~\cite{kfir2019entanglements,di2019probing}.
In particular, the vacuum spontaneous processes exhibit intriguing prospects for universal and ultrafast quantum computation and information processing because they can create a variety of highly non-trivial quantum optical states under designed interaction sequences~\cite{dahan2023creation,baranes2023free} and they are essential for the proposed hybrid free-electron-polariton blockade~\cite{PRXQuantum.5.010339}.

The key to these predictions is a strong quantum interaction with $\abs{\gqu}>1$.
Intuitively, this regime indicates that each emitted photon can cycle back to the electron and cause a cascade of multi-photon emission and absorption. 
Crucially, this strong interaction regime involves higher-order quantum electrodynamics processes and can be understood as inducing a unique type of optical nonlinearity. 
Strong optical nonlinearities are being pursued in all fields of photonics, with the goal of reaching single-photon nonlinearities. 
Thus, a question of fundamental importance is whether electron-photon interactions can reach this regime. 
Answering this question requires an upper bound to the quantum interaction strength $\gqu$ in arbitrary photonic environments.

In this work, we derive universal limits to the quantum interaction strength $\gqu$ between free electrons and single-mode photons and provide analytical and numerical validations.
We find that the limits allow for the strong quantum interaction condition $\abs{\gqu}>1$, which is feasible in near-optimal designs under realistic conditions.
Limits for both line and point electrons are obtained for arbitrary geometries, relying only on material electrostatic constants, electron velocity, and electron-structure separation.
Intriguingly, the limits produce an explicit criterion for selecting electron and photon energy under certain electron-structure separations and give rise to two optimal interaction regimes, one favoring fast electrons and the other favoring slow electrons.
We evaluate the limits across the electromagnetic spectrum and reveal the dichotomy of fast and slow electrons: the former is advantageous for high-energy photons, while the latter predominates towards the long-wavelength regime.

%--------------------------------------
\begin{figure}[htbp]
	\centering
	\includegraphics[width=1\linewidth]{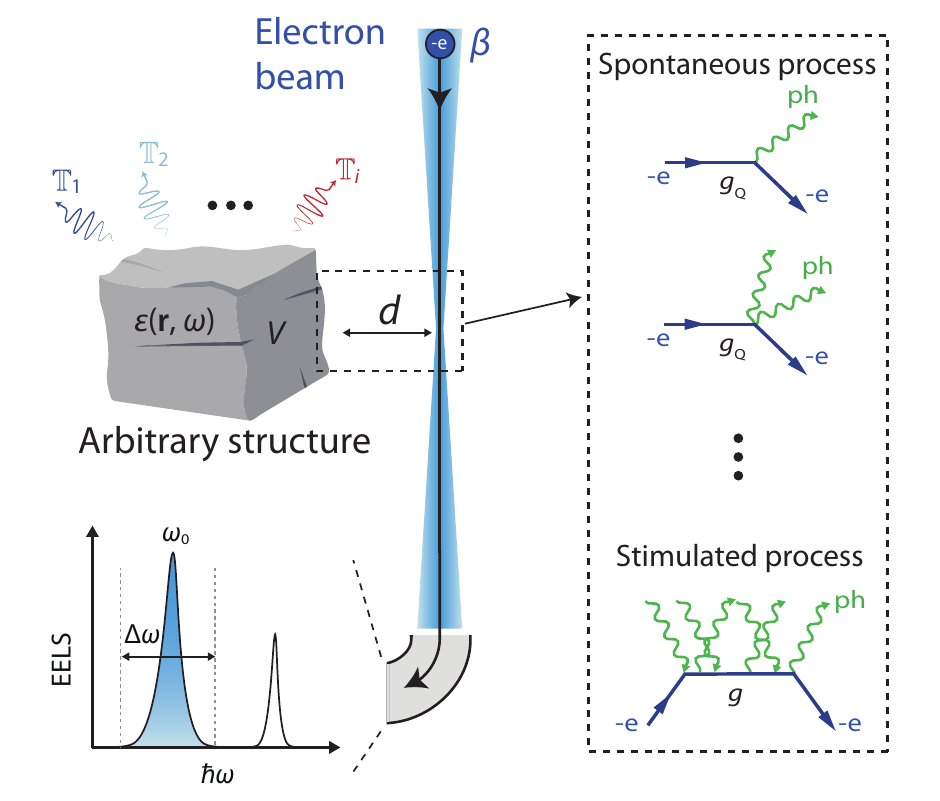}
	 \caption{%
	 	\textbf{Illustration of the theoretical framework.} 
	 	Schematic of an electron beam with velocity $\beta = v/c$ traveling close to an arbitrary structure at a separated distance $d$. The structure is defined by material permittivity $\epsilon(\rv,\omega)$ within a volume $V$. The electron (e$^-$) can become an entangled state with its emitted photons (ph) through spontaneous or stimulated processes. These quantized emission events are able to be identified via EELS and PINEM, respectively. The inset Feynman diagrams show the corresponding physical processes and represent the interaction between electrons and photons can be described by a spontaneous quantum strength $\gqu$ in EELS and a stimulated quantum strength $\mathit{g}$ in PINEM. The spontaneous quantum interaction strength $\gqu$ relates to the electron-energy-loss spectral density by $\lvert \gqu \lvert^2$ = $\int \Gamma(\omega)\,\mathrm{d} \omega$, where the integration window $\Delta \omega$ relates to the intended mode(s) that free electrons couple to. The aforementioned interaction can be described by a special scattering matrix “$\Tmatrix$”, which can be decomposed into a set of Drude–Lorentz oscillators with matrix-valued coefficients $\Ti$.
        }
	\label{fig_main: schematic}
\end{figure}
%--------------------------------------

We begin the analysis by considering an electron beam at normalized velocity $\beta = v/c$, interacting with an arbitrary photonic structure characterized by volume $V$ and material permittivity $\epsilon(\rv,\omega)$ in its near field with a separation $d$, as shown in Fig.~\ref{fig_main: schematic}. 
Photon exchanges between the electron and the structure can be strongly amplified when the photonic structure supports modes with large spatial and spectral overlap with the electron’s excitation field.
In the framework of quantum optics, the joint electron (e$^{-}$)-photon (ph) state $\ket{\Psi}=\ket{\psi^{\mathrm{e}}} \otimes \ket{\psi^{\mathrm{ph}}} =\sum_{n=0}^{\infty}\sum_{k=-\infty}^{+\infty}c_{n, k}\ket{\mathit{E}_{k}, n}$ is a superposition state with electron energy $\mathit{E}_0 + \mathit{k} \hbar \omega$ and $\mathit{n}$ photons where $n$ denotes the photonic Fock states and $k$ denotes the electronic energy ladder.
For spontaneous processes with an electron of energy $\textit{E}_0$ in an empty cavity, the initial state is $\ket{\Psi_{\mathrm{i}}} = \ket{\textit{E}_0,0}$, and the final state $\ket{\Psi_{\mathrm{f}}}$ satisfies $n=-k$ by virtue of energy conservation. 
The initial and final states are related by the scattering matrix, represented as $\ket{\Psi_{\rm{i}}} =  \smatrix \ket{\Psi_{\rm{f}}}$, which is given by~\cite{di2019probing,kfir2019entanglements}
\begin{equation}
    \smatrix=\exp\left(\gqu \hat{\mathit{b}}^{\dagger} \hat{\mathit{a}} -\gqu^* \hat{\mathit{b}} \hat{\mathit{a}}^\dagger  \right),
    \label{eq:S-matrix}
\end{equation}
where $\hat{\mathit{a}}$, $\hat{\mathit{a}}^{\dagger}$ are the annihilation and creation operators of the optical mode; $\hat{\mathit{b}}$, $\hat{\mathit{b}}^{\dagger}$ denote the ladder operators for electron energy, describing energy exchanges between the electron and the optical mode; and $\gqu$ is the dimensionless electron-photon interaction strength.
The strong-coupling regime of $\abs{\gqu}>1$ indicates that one free electron becomes more likely to emit more than one photon and that this electron becomes quantum entangled with these photons.
In the limit of vacuum zero field~\cite{garcia2021manipulating}, the quantum-optical theory of PINEM coincides with the theory of electron energy loss spectroscopy (EELS), which enables the calculation of $\gqu$ from the spectrally-integrated electron energy loss probability (see SI Sec. S1):  $\lvert \gqu \lvert^2=\int \Gamma(\omega) \,\mathrm{d}\omega$ without competing loss mechanisms~\cite{kfir2021optical,di2021modulation,huang2023electron,baranes2023free}.
In stimulated processes like PINEM, the general quantum framework still holds, whereas one should replace $\gqu$ with the stimulated strength $g$ amplified by the input coherent state amplitude.

To impose a bound on $\gqu$ and $g$, we treat free electrons interacting with an arbitrary photonic environment as an electromagnetic scattering problem (Fig.~\ref{fig_main: schematic}).
Although this scattering perspective has previously led to a single-frequency (zero-bandwidth) limit~\cite{yang2018maximal}, it cannot translate to a $\gqu$ limit straightforwardly because of a divergent material metric for lossless materials. 
Alternatively, a sum-rule approach can bound near-field interactions (\eg power loss
from a stationary point dipole)~\cite{shim2019fundamental,Miller2023}, but its applicability to free electrons is hindered by the non-analytic behaviors of their external fields near the complex-plane origin.
Recently, a $\Tmatrix$-matrix--based oscillator representation framework~\cite{zhang2023all} (where the $\Tmatrix$ operator relates the polarization fields with incident fields~\cite{carminati2021principles,chao2022physical}) appears ideal for analyzing free-electron--light scattering.

In the polarization-response representation, the frequency-dependent extinction equals the work done by the incident field on the induced polarization field in the scatterer, $\Powerext(\omega)  = \left({\omega}/{2} \right) \mathrm{Im}\int_{\textit{V}}\Einc^{*}\cdot\mathbf{P}\,\mathrm{d}\mathit{V}$. 
The polarization response is related to the incident field through a linear operator, the ``$\Tmatrix$ matrix"~\cite{carminati2021principles}:  $\polfield(\mathbf{r}) = \int_V \Tmatrix(\mathbf{r},\mathbf{r^{\prime}}) \cdot \Einc( \mathbf{r^{\prime}} ) \, \mathrm{d} \mathbf{r^{\prime}} $, or $\mathbf{p} = \Tmatrix\einc$ in vector-matrix notation that we use throughout.
Causality and passivity produce a Kramers–Kronig relation and sum-rule constraints for the $\Tmatrix$ matrix. 
These properties enable the decomposition of the $\Tmatrix$ matrix into an infinite sum of lossless Drude–Lorentz oscillators with resonance frequency $\omega_i$ and highly constrained matrix-valued oscillator strengths $\Ti$ representing the only degrees of freedom~\cite{zhang2023all} (see Sec. S2). 
Contributions from $\Ti$ at well-separated $\omega_{i}$ originate from different resonances.
For reciprocal systems, combining the zero-frequency sum rule, the oscillator representation, and the domain monotonicity of polarizability~\cite{shim2019fundamental} along the positive-frequency axis yields $\omega\mathrm{Im}\Tmatrix(\omega)
\leq \left( {\pi \econstant}/{2} \right) \sum_{i=1}^{\infty} \omega^2 \Tmatrix_i \delta(\omega-\omega_i)$, which can reformulate the extinction as 
\begin{equation}
\begin{aligned}
\Powerext(\omega) 
&\leq \frac{\omega^2\pi \epsilon_0 \econstant}{4} \sum_{\textit{i}=1}^{\infty} \einc^\dagger \Tmatrix_\textit{i} \einc \delta(\omega-\omega_\textit{i}).
\label{eq:T2ext}
\end{aligned}
\end{equation}
The constant $\econstant$ is a dimensionless electrostatic permittivity coefficient relating the incident and total fields at zero frequency via $\mathbf{P}(0)=\epsilon_0\tau\Einc(0)$ ($\epsilon_0$ is the vacuum permittivity).

Based on this foundation, we next derive the upper limit to the vacuum strength $\gqu$ of free-electron-photon interaction. 
Under certain interaction bandwidth described by a dimensionless window function $\Theta(\omega; \omegam,\Delta\omega)\leq 1$ (peaked at $\omegam$ with bandwidth $\Delta\omega$), $\gqu$ can be evaluated by integrating the loss probability $\Gamma(\omega) = \Powerext(\omega)/\hbar\omega$: $\abs{\gqu}^2 = \int \Gamma(\omega) \Theta(\omega; \omegam,\Delta\omega) \, \mathrm{d} \omega$.
By combining Eq.~\eqref{eq:T2ext}, $\abs{\gqu}^2 $ becomes
\begin{equation}
\abs{\gqu}^2
\leq \frac{\pi \epsilon_0 \econstant}{4 \hbar} \mathrm{Tr} \sum_{i=1}^{\infty} \Tmatrix_i \Bi,
\label{eq:int-ext}
\end{equation}
a $\Ti$-weighted sum of $\Bi \equiv$~$\omega_i \einc(\omega_i) \einc^\dagger(\omega_i) \Theta(\omega_i;{\omegam,\Delta\omega})$ that incorporates the frequency of the emitted photons, the incident field, and the window function. 
Because $\omega \einc(\omega) \einc^\dagger(\omega)$ varies much slower than $\Theta(\omega; \omegam,\Delta\omega) $ under a high-Q interaction scenario ($\omega_0\gg\Delta\omega$), $\Bi$ also spectrally peaks at $\omega_0$ with negligible detuning.
Therefore, setting $\Ti=\mathbb{I}_V$ for the oscillator at $\omega_0$ and zero elsewhere will maximize Eq.~\eqref{eq:int-ext}, which simultaneously satisfies the sum-rule constraint $\sum_{i=1}^{\infty} \Tmatrix_i = \mathbb{I}_{V}$, where $\mathbb{I}_{V}$ is an identity matrix within the scatterer volume $V$.
This mathematically optimal choice corresponds to the physical condition of all of the sum-rule-constrained polarization responses concentrating within the frequency window of interest.

Applying the single-mode condition at $\omega_0$ to Eq.~\eqref{eq:int-ext}, we get
\begin{equation}
\abs{\gqu}^2 \leq 
\frac{\pi \epsilon_0 \omegam \econstant}{4 \hbar} \int_V \abs{ \Einc(\omegam)}^2 \,\mathrm{d} V .
\label{eq:bound}
\end{equation}
This limit to $\gqu$ thus becomes a simple product among physical constants, photon frequency, electrostatic polarizability, and the integration of the incident fields within the scatterer.  
We can obtain $\econstant$ analytically for free electrons with canonical geometries (\eg $\econstant = 2\epsilon_1 (\epsilon_2-1) / (\epsilon_2+\epsilon_1)$ for a half-space, and $\econstant = \epsilon_1 (\epsilon_2 - 1)/{\epsilon_2}$ for a concentric cylinder, both made of materials $\epsilon_1$ and $\epsilon_2$, where $\epsilon_1$ and $\epsilon_2$ are electrostatic permittivity of materials 1 and 2, respectively; see Sec. S2.D).
Evidently, $\tau$ is finite and bounded for both dielectrics and perfect conductors (see Sec. S2.D).

%--------------------------------------
\begin{figure}[htbp]
	\centering
	\includegraphics[width=\linewidth]{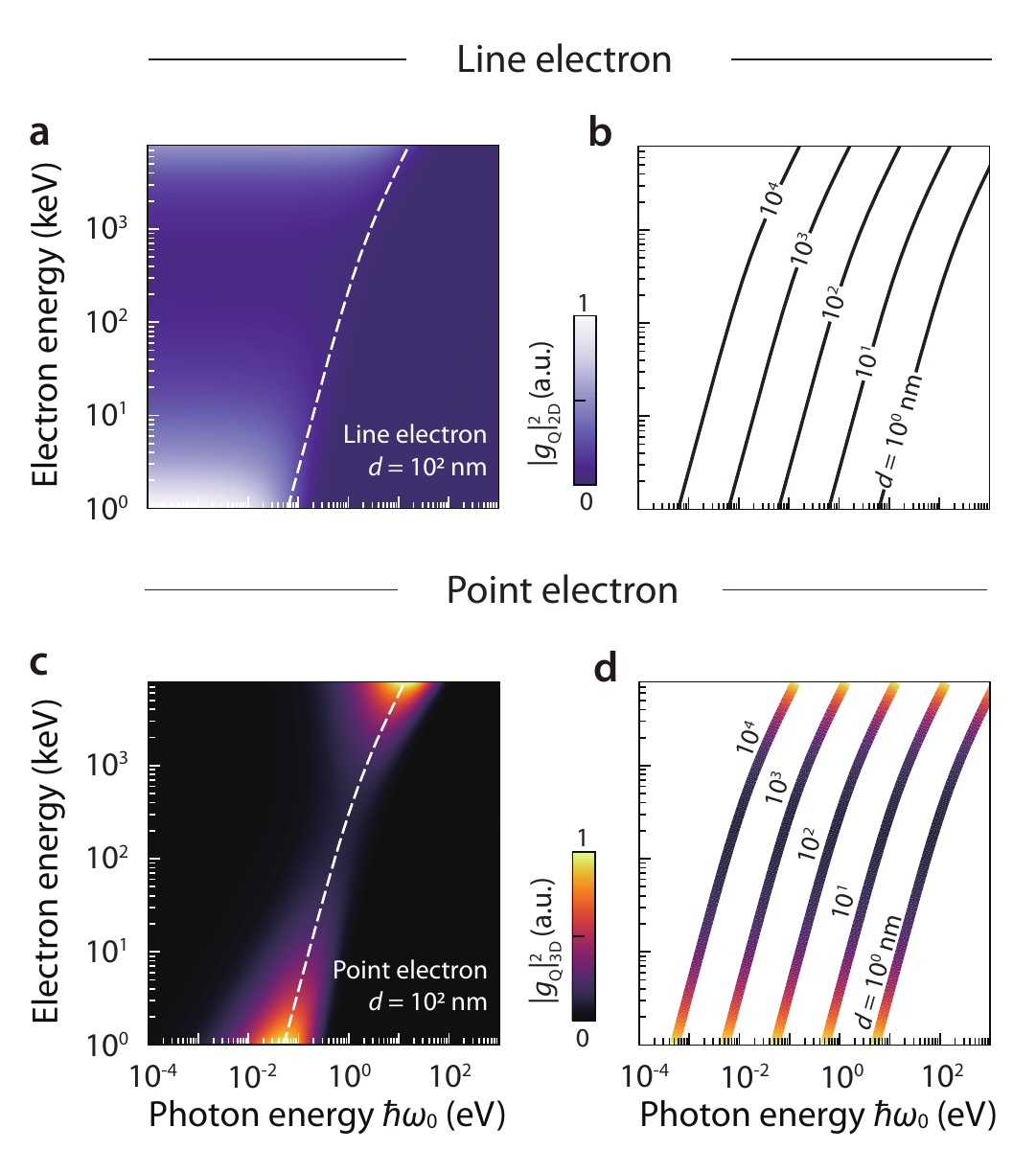}
	 \caption{%
	 	\textbf{Universal behaviors of the quantum interaction strength $\abs{\gqu}^2$.}
            \textbf{a.} Line-electron upper limit Eq.~\eqref{eq:2D-limit_main} as a function of electron energy and photon energy under a separation distance $d = \SI{100}{\nm}$. 
            The quantum interaction strength is normalized between 0 and 1 by the maximum $\gqusheet$, and the white dashed line denotes the cut-off photon energy of the optimal region.
            \textbf{b.} Photon energy cut-off of line-electron limit Eq.~\eqref{eq:maintext2D-cutoff} across various separation $d$.
            \textbf{c.} Point-electron limit Eq.~\eqref{eq:3D-limit_main} for $d =~\SI{100}{\nm}$.
            The white dashed line highlights optimal choices of electron and photon energies.
            \textbf{d.} The optimal electron and photon energies under different separation $d$ [Eqs.\eqref{eq:maintext3D-cutoff} and \eqref{eq:4064}]. 
	    }
	\label{fig_main: predictions}
\end{figure}
%--------------------------------------

%--------------------------------------
\begin{figure*}[htbp]
	\centering
	\includegraphics[width=0.89\linewidth]{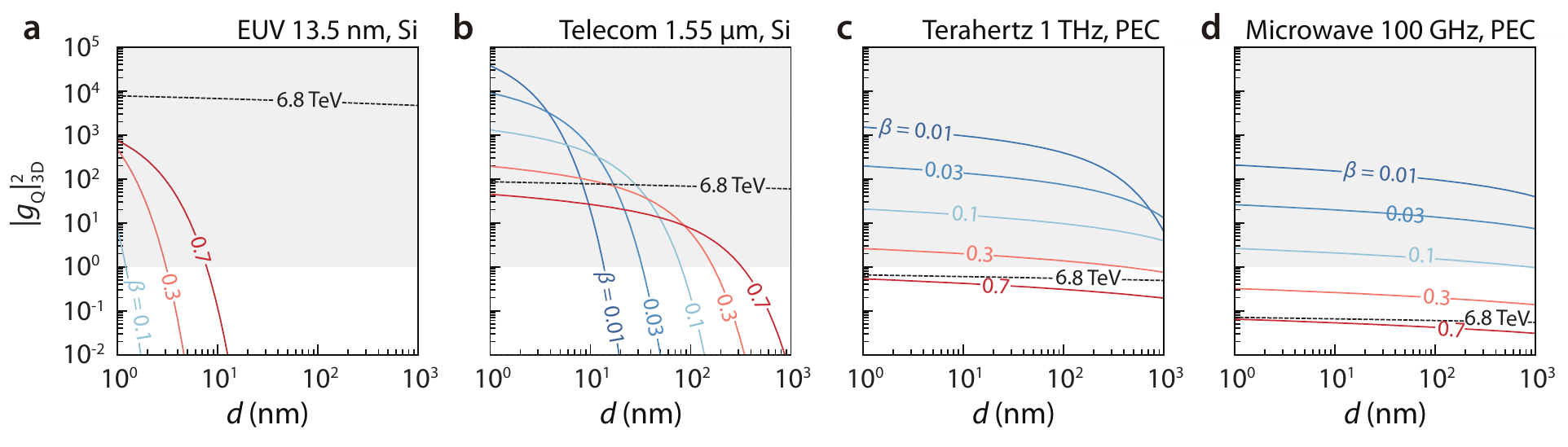}
	 \caption{%
	 	\textbf{$\abs{\gqu}^2$ upper limit at typical photon frequencies across the electromagnetic spectrum.}
	 	\textbf{a.} Silicon at extreme ultraviolet (EUV)~\SI{13.5}{\nm}.
            \textbf{b.} Silicon at telecommunication~\SI{1.55}{\um}.
            \textbf{c.} Perfect electric conductor (PEC) at~\SI{1}{\THz}.
            \textbf{d.} PEC at microwave~\SI{100}{\GHz}. 
        We vary electron velocities $\beta$ and fix the interaction length at~\SI{100}{\um}.
        The black dashed line represents ultra-relativistic \SI{6.8}{TeV} electrons.
        The light gray shading indicates the $\abs{\gqu}^2 > 1$ regime.
	 	}
	\label{fig_main: emission center frequency}
\end{figure*}
%--------------------------------------

We further obtain shape-independent upper limits for line and point electrons by enclosing arbitrary scatterers with canonical geometries.
Assuming translational invariance along one of its transverse directions and the spatial coherence of the source current~\cite{remez2019observing}, a line electron beam (sometimes also denoted as ``sheet electron beam") is a simplified mathematical treatment used in many free-electron theory analyses~\cite{luo2003cerenkov,andrews2004gain,kumar2006analysis,wang2016manipulating,pan2023polarizing}.

Applying Eq.~\eqref{eq:bound} to a line electron beam moving above a half-space (which can enclose an arbitrary scatterer) at a separation $\textit{d}$ yields a shape-independent limit for 2D line electrons (see Sec. S3):
\begin{equation}
\gqusheet  \leq 
\frac{\pi \chargedensity^2 \tau L}{32 \hbar \epsilon_0}  \frac{ \left( \kvmax^2 + \kapparhomax^2 \right) }{\omegam\kapparhomax} \mathrm{e}^{-2 \kapparhomax d},
\label{eq:2D-limit_main}
\end{equation}
where $\chargedensity$ is the charge density per unit transverse length, $L$ is the interaction length,  and $\kmax = \omegam/c$, $\kvmax = \omegam/v$,  $\kapparhomax = \sqrt{\kvmax^2 - \kmax^2} = \kmax/\beta \gamma $ are the free-space, electron longitudinal and transverse wavevectors of the target photon frequency $\omegam$, respectively and $\gamma = 1/ \sqrt{1-\beta^2}$ is the Lorentz factor.

Applying Eq.~\eqref{eq:bound} to a point electron moving at the center of a concentric hollow cylinder sector (which can enclose an arbitrary scatterer) of azimuthal opening angle $\psi$, inner radius $d$, and outer radius $\infty$ leads to a shape-independent limit for point electrons (see Sec. S4):
\begin{equation}
\gqupoint \leq
\frac{\alpha\econstant \psi L }{4c}\frac{ \omegam }{\beta^2} \kapparhomax d \BesselKzero(\kapparhomax d) \BesselKfirst(\kapparhomax d),
\label{eq:3D-limit_main}
\end{equation}
where $\alpha$ is the fine-structure constant, $K_{\mathrm{n}}$ is the modified Bessel function of the second kind.
The limits Eqs.~\eqref{eq:bound}-\eqref{eq:3D-limit_main} are the key analytical findings of the paper.
They show that the maximum $\abs{\gqu}$ for a target emission center frequency $\omegam$ is solely determined by the electron velocity (via $\beta$ and $\kapparhomax$), the separation $d$, and the electrostatic polarizability $\tau$ of the optical environment. 

We illustrate the universal behaviors of the quantum upper limits in Fig.~\ref{fig_main: predictions} to depict the key features of Eqs.~\eqref{eq:2D-limit_main} and \eqref{eq:3D-limit_main}, ignoring the fractional pre-factors comprising physical constants and structural coefficients. 
This way, we can obtain a general energy selection recipe under arbitrary separations.

For line electrons (Fig.~\ref{fig_main: predictions}a), a photon energy cut-off condition (dashed line in Fig.~\ref{fig_main: predictions}a; 
Sec. S5) appears:
\begin{equation}
\kapparhomax d  = \omegam d/ c \beta \gamma = 0.5,
\label{eq:maintext2D-cutoff}
\end{equation}
meaning that large $\abs{\gqu}$ is possible and forbidden below and above the cut-off photon energies, respectively, as indicated by the dichotomy of the brightness and darkness to the left and right of the cut-off line in Fig.~\ref{fig_main: predictions}a.
As the separation distance reduces, the cut-off line blueshifts in photon energy (Fig.~\ref{fig_main: predictions}b).

For point electrons (Fig.~\ref{fig_main: predictions}c), an optimal condition (dashed line in Fig.~\ref{fig_main: predictions}c and see Sec. S6) illuminates the choice of electron and photon energies to maximize their interaction:
\begin{equation}
\BesselKzero(\kapparhomax d)\BesselKfirst(\kapparhomax d) = \kapparhomax d \left[ \BesselKzero(\kapparhomax d)^2 +\BesselKfirst(\kapparhomax d)^2 \right].
\label{eq:maintext3D-cutoff}
\end{equation}
Such an optimal condition is satisfied when 
\begin{equation}
    \kapparhomax d\approx0.4064,
    \label{eq:4064}
\end{equation}
which differs from line electrons' cut-off condition Eq.~\eqref{eq:maintext2D-cutoff}.  
In particular, two optimal regimes emerge and favor relativistic (top-right brightness in Fig.~\ref{fig_main: predictions}c, ultraviolet~\SI{10}{\eV} photons coupled to $\gtrapprox\SI{6}{\MeV}$ electrons when $d=\SI{100}{\nm}$) and non-relativistic (bottom-left brightness in Fig.~\ref{fig_main: predictions}c, long-infrared \SI{0.1}{\eV} photons coupled to $\lessapprox\SI{4}{\keV}$ electrons when $d=\SI{100}{\nm}$) electrons, respectively.
The former optimal regime aligns with traditional free-electron physics, where high-energy photons are generated with relativistic electrons, whereas the latter optimal regime indicates the huge potential for quantum electron-light interaction with slow electrons, which has been recognized in an early study on multi-plasmon generation in graphene~\cite{garciía2013multiple} and rapid development is being made in more recent years~\cite{liu2017integrated,yang2018maximal,talebi2020strong,liebtrau2021spontaneous,huang2023quantum,karnieli2023jaynes,chen2023low,pan2023polarizing,eldar2024self,shiloh2022quantum,rasmussen2024generation,karnieli2024strong,giulio2024toward}.
For intermediate electron energies, the optimal photon energies can be analogously defined using Eq.~\eqref{eq:maintext3D-cutoff}, although the associated limit is lower than those for the two optima. 
The two optimal regimes prevail under various separation distances (Fig.~\ref{fig_main: predictions}d).

We next contextualize the discussion by applying the universal feature above to evaluate the feasibility of $\abs{\gqu}>1$  at concrete frequency windows.
Under various electron velocities $\beta$ and separation $d$, Fig.~\ref{fig_main: emission center frequency} shows the $\abs{\gqu}^2$ upper limit for point electrons [Eq.~\eqref{eq:3D-limit_main}] in four technologically relevant regimes, namely, extreme ultraviolet (EUV) (Fig.~\ref{fig_main: emission center frequency}a), near-infrared (Fig.~\ref{fig_main: emission center frequency}b), Terahertz (Fig.~\ref{fig_main: emission center frequency}c), and microwave (Fig.~\ref{fig_main: emission center frequency}d).
% 13.5 nm
At the EUV \SI{13.5}{\nm} (Fig.~\ref{fig_main: emission center frequency}a), despite the high limits near the atomic-scale separations (under which quantum tunneling and surface effects can become pronounced and modify the bounds here~\cite{yang2019general,gonccalves2023interrogating,zhu2016quantum}), $\abs{\gqu}>1$ is almost impossible for $d\gtrapprox\SI{10}{\nm}$ except with $\beta\gtrapprox0.8$ relativistic electrons. 
%1.55 um
At the telecom \SI{1.55}{\um} (Fig.~\ref{fig_main: emission center frequency}b), whereas slow electrons' upper limits are higher at tens-of-nanometer or smaller separation, fast electrons with $\beta\gtrapprox0.3$ are superior at $d\gtrapprox\SI{100}{\nm}$, a typical separation in modern grazing-interaction experiments.
Towards the longer wavelengths, the advantages of slow electrons are predominant. $\abs{\gqu}>1$ is only possible for $\beta\lessapprox0.3$ and $\beta\lessapprox0.1$ at \SI{1}{\THz} (Fig.~\ref{fig_main: emission center frequency}c) and \SI{100}{\GHz} (Fig.~\ref{fig_main: emission center frequency}d), respectively.
Despite the $\ln(\gamma)$ divergence of the limit Eq.~\eqref{eq:3D-limit_main} in electron energy, the slow-electron advantage remains evident when comparing with electrons of \SI{6.8}{\TeV} (dashed lines in Fig.~\ref{fig_main: emission center frequency}c and d), the current highest acceleration energy~\cite{ciccotelli2023energy}.
Taken together, the $\abs{\gqu}>1$ goal is indeed challenging for EUV and X-ray photons; nevertheless, slow electrons are promising from electrostatics to the far-infrared, and fast electrons are advantageous from mid-infrared to far ultraviolet; the evaluation is based on the current levels of electron beam collimation and focusing.

%--------------------------------------
\begin{figure*}[htbp]
	\centering
	\includegraphics[width=0.785\linewidth]{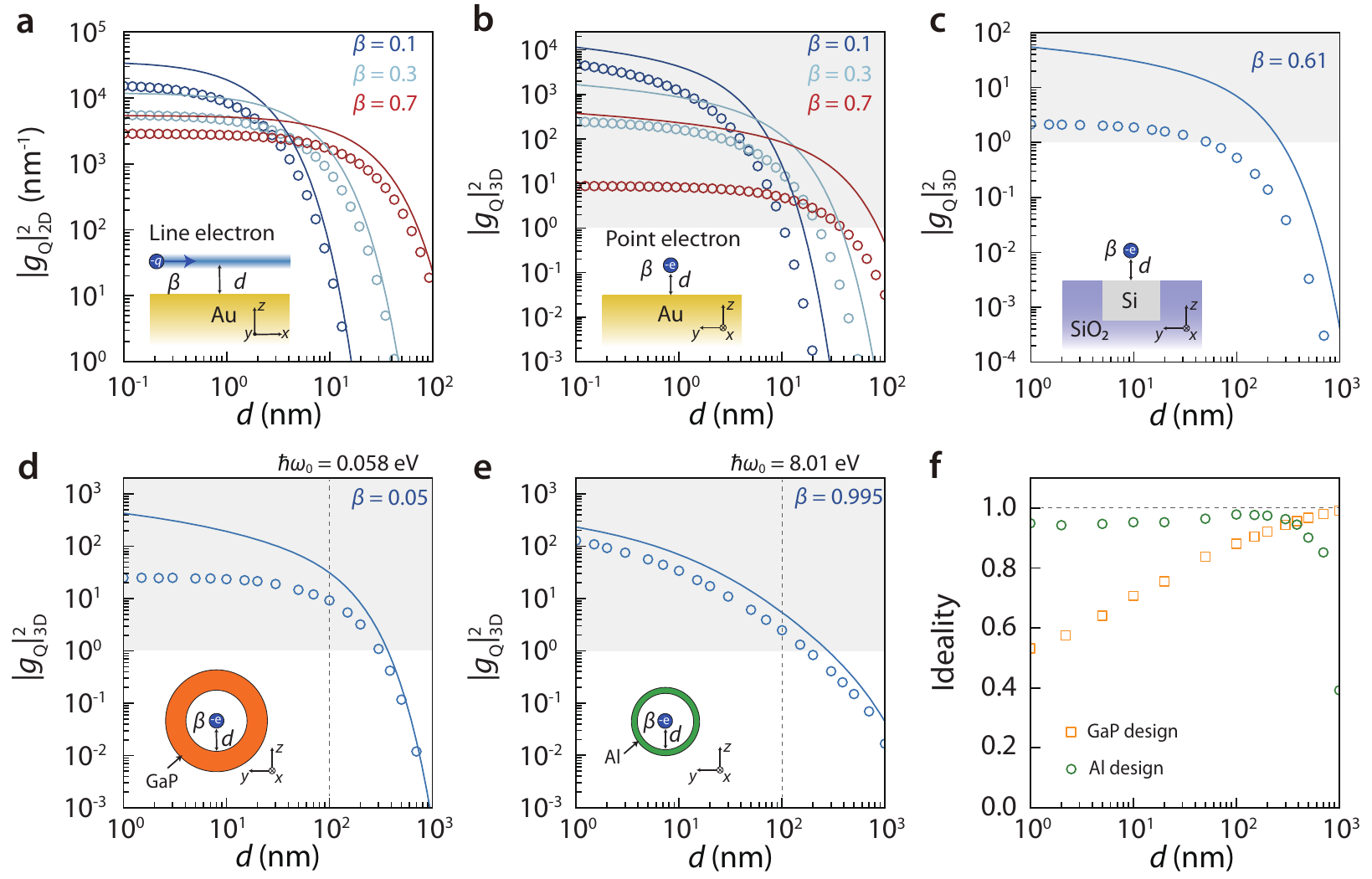}
	 \caption{%
	 	\textbf{Validations (a-c) of the theoretical upper limits and near-optimal designs (d-f).}
            \textbf{a-b.} Analytical calculations compared to the upper limits.
            A line electron~(\textbf{a}) with charge density $\chargedensity$ = $1.6 \times 10^{-19}~\rm{C/nm}$ and a point electron~(\textbf{b}) interacting with a half space made of air and a Drude metal ($\epsilon_{\rm{m}} = 1 - \omegap^2 / \omega(\omega + \iu \gamma_{\rm{m}})$, $\omegap = \SI{9.06}{\eV}$, $\gamma_{\rm{m}} = \SI{0.071}{\eV}$~\cite{echarri2021optical} for gold).
	 	\textbf{c.} Numerical simulations compared to the upper limits. A point electron moves above a Si-SiO$_2$ waveguide (thickness~\SI{220}{\nm} and width~\SI{500}{\nm}) at a distance $d$.
            \textbf{d.} Near-optimal electron-photon coupling design for slow electrons.
           An air-GaP (Gallium Phosphide) core-shell structure with a~\SI{40}{\nm}  GaP shell is designed for non-relativistic electrons.
            The property of polar dielectric GaP is described by a Lorentz oscillator model, $\epsilon_{\mathrm{d}} = \epsilon_\infty + 
            \epsilon_\infty ( \omega_{\mathrm{LO}}^2 - \omega_{\mathrm{TO}}^2 ) / (\omega_{\mathrm{TO}}^2 - \omega^2 - \iu \omega \gamma_{\mathrm{d}} )$, where background permittivity $\epsilon_\infty =~\SI{9.1}{}$~\cite{van1969quantum}, $\gamma_{\mathrm{d}} =~\SI{1.25}{\meV}$, longitudinal optical $\omega_{\mathrm{LO}}$ and transverse optical $\omega_{\mathrm{TO}}$ phonon frequencies are~\SI{67.8}{\meV} and~\SI{47.1}{\meV}, respectively~\cite{caldwell2015low}.
            \textbf{e.} Near-optimal electron-photon coupling design for fast electrons.
            An air-aluminum core-shell structure with a~\SI{13}{\nm} aluminum (Al)~\cite{rakic1995algorithm} shell is used for relativistic electrons.
            The light gray shows the $\abs{\gqu}>1$ regime (circles: exact calculations; lines: theoretical upper limit).   
            \textbf{f.} The coupling ideality as a function of separation (orange squares: GaP design; green circles: Al design). 
            The interaction length is fixed at~\SI{100}{\um} for all calculations.
            }

	\label{fig_main:validations}
\end{figure*}
%--------------------------------------

We analytically and numerically validate the derived upper limits in Fig.~\ref{fig_main:validations}a-c.
We first consider two analytical scenarios: line (Fig.~\ref{fig_main:validations}a) and point (Fig.~\ref{fig_main:validations}b) electron beams pass above a Drude-metal half-space in vacuum and launch surface plasmon polaritons (SPPs).
Both cases permit analytical treatment of energy loss (see derivations in Sec. S7) to validate the line-electron [Eq.~\eqref{eq:2D-limit_main}] and point-electron [Eq.~\eqref{eq:3D-limit_main}] upper limits, respectively.
We compute $\gqu$ when electrons excite SPPs at frequency $\omega_{\mathrm{SPP}}$ that vary under different electron velocities.
The definite integral of energy loss is calculated between ($\omega_{\rm{SPP}}$ - $\gamma_{\mathrm{m}}$) and ($\omega_{\rm{SPP}}$ + $\gamma_{\mathrm{m}}$) for an interval of two full-width-half-maximum, where $\gamma_{\mathrm{m}}$ is simultaneously the intrinsic material loss and the surface plasmon decay rate of the Drude metal (see Sec. S8). 
In both the 2D and 3D cases, the analytical results closely trail the upper limits at the considered separation.
% numerical validation 
We also numerically validate the upper limit using full-wave simulations of an integrated waveguide setup in Fig.~\ref{fig_main:validations}c, where an electron passes above a $\rm{Si}$ waveguide embedded in $\rm{SiO_2}$ substrate; 
the numerical coupling strength $\gqu$ of the fundamental mode of the waveguide falls short of the upper limit for about one order of magnitude.

Finally, we design two near-optimal hollow-core structures (Fig.~\ref{fig_main:validations}d-f) to verify the two predicted optimal regimes (two bright areas in Fig.~\ref{fig_main: predictions}b) associated with slow and fast electrons, respectively.
The core-shell configuration is motivated by the sum-rule constraint constant $\tau$, which is maximum when the largest possible near-field area of a source is occupied by the polarizable material.
For the slow-electron optimum, we utilize a Lorentz polar dielectric gallium phosphide (GaP) with low optical phonon frequencies to allow the coupling between \SI{0.058}{\eV} infrared photons and non-relativistic $\beta=0.05$ electrons;
for the fast-electron optimum, we adopt Aluminium that features a high plasma frequency to achieve an efficient interaction between \SI{8.01}{\eV} ultraviolet photons and $\beta=0.995$ relativistic electrons. 
In both cases, $\abs{\gqu}>1$ is predicted possible by the upper limits and is validated by the slightly trailing concrete designs at the separation \SI{100}{\nm} (dashed vertical lines in Fig.~\ref{fig_main:validations}d-e) and interaction length \SI{100}{\um} that are relevant to current electron beam focusing and collimation techniques.
In Fig.~\ref{fig_main:validations}d and e, GaP and Aluminium, two lossy materials, are used primarily to verify the two optimal regimes (Fig.~\ref{fig_main: predictions}c), as their material properties facilitate the coupling of electrons and photons at the desired energy. 
Alternatively, lossless platforms (\eg Fig.~\ref{fig_main:validations}c) can be employed to maintain the entanglement of the interaction.

In addition, Fig.~\ref{fig_main:validations}f depicts the coupling ideality of the two near-optimal designs as the function of separation. 
The ideality of the target mode \textit{m} is defined as $I_m = \abs{ \mathit{g}_{\rm{Qu, }\mathit{m}}} ^2 / \sum_{\textit{i}} \abs{ \mathit{g}_{\rm{Qu, }\mathit{i}} } ^2$, which describes the fraction of coupling into the designated mode among exciations of all modes~\cite{di2019probing, huang2023electron}.
The ideality of the slow-electron design (Fig.~\ref{fig_main:validations}d; orange squares in Fig.~\ref{fig_main:validations}f) grows monotonically with the considered separation, reaching more than 95\% under $\textit{d} >~\SI{500}{\nm}$ due to the dominance of the fundamental mode.
Meanwhile, the fast-electron design (Fig.~\ref{fig_main:validations}e; green circles in Fig.~\ref{fig_main:validations}f) exhibits a good ideality $>80\%$ for a wide range of separations, including both near- and far-field of the \SI{8.01}{\eV} photon.

In summary, we have theoretically derived a universal upper limit to the quantum interaction strength $\gqu$ between free electrons and single-mode photons.
The limit allows us to evaluate the feasibility of achieving the $\abs{\gqu}>1$ strong interaction condition
over all possible designs across the electromagnetic spectrum without exhaustive computational optimization.
Under arbitrary separations, the limits identify two optimal interaction regimes, showing how to select electron and photon energy to maximize their interaction.
Notably, the $\gqu$ limit developed here can directly translate to a limit for the stimulated strength $g$, scaled by the coherent state amplitude.
There still exists some gap between the limits and the performance of prevalent structures, indicating the potential for further tightening the limits and improving designs to approach them.

The limits derived here hinge on reciprocity, under which the skew-symmetric part of the $\Tmatrix$-matrix vanishes, and the sum rule of the real-symmetric part of $\Tmatrix$ can be applied; it is thus pertinent to seek the generalization of the current limits to nonreciprocal free-electron systems~\cite{yu2021inelastic,prudencio2018asymmetric,fallah2021nonreciprocal} and explore the possibility for higher $\gqu$ therein.
As the electron-structure separation increases into the far field, electron energy loss undergoes exponential decay, and its spectral composition can become multimodal. This scenario may violate the single-mode condition, decrease the coupling ideality, and cause an apparent breakdown of the limits developed here because the spectral integration of energy loss no longer corresponds to the $\abs{\gqu}^2$ of a particular mode of interest (\eg when the tail of the zero-loss peak dominates; see Sec. S9).
Therefore, it would also be of interest to generalize the present work into the multimodal condition.
The findings here offer an intuitive understanding of how to maximize the quantum interaction between free electrons and photons.
They also provide practical design guidelines for future experiments that aim to achieve strong entanglement between electrons and photons, as well as photon entanglement mediated by free electrons.

\emph{Acknowledgments.}
We thank J.~Garc\'{i}a~de~Abajo, \mbox{F.~J.~Garc\'{i}a-Vidal}, Yixin~Sha, and Zhexin~Zhao for fruitful discussions. 

\textit{Note added.} During the completion of this manuscript, we became aware of related work~\cite{zhao2024arxiv}.

%%%%%%%%%%%%%%%%%%%%%%%%%%%%%%%%%%%%%%%%%%%%%%%%%%%%%%%%%%%%%%%%%%%%%%%%%%%%%%%%%%%%%%%%%%%%%%


\end{document}